\documentclass{elsarticle}
\usepackage{lineno,hyperref}
\usepackage{placeins, graphicx}
\modulolinenumbers[5]

\begin{document}

\begin{frontmatter}
\title{Mass Segregation Phenomena using the Hamiltonian Mean Field Model}

\author{J.R. Steiner\textsuperscript{1}}
\author{Zolacir T.O.Jr.\textsuperscript{1,2}}
\address{Departamento de Ciencias Exatas e Tecnologicas, Universidade Estadual de Santa Cruz, Ilheus, 45662 900, Bahia, Brazil}
\address{Instituto de F\'isica and International Center for Condensed Matter Physics Universidade de Bras\'ilia, CP: 04455, 70919-970 - Bras\'ilia, Brazil}
\bigskip
\date{}
\begin{abstract}
Mass segregation problem is thought to be entangled with the dynamical evolution of young stellar clusters \cite{olczak}. This is a common sense in the astrophysical community. In this work, the Hamiltonian Mean Field (HMF) model with different masses is studied. A mass segregation phenomenon (MSP) arises from this study as a dynamical feature. The MSP in the HMF model is a consequence of the Landau damping (LD) and it appears in systems that the interactions belongs to a long range regime. Actually HMF is a toy model known to show up the main characteristics of astrophysical systems due to the mean field character of the potential and for different masses, as stellar and galaxies clusters, also exhibits MSP. It is in this sense that computational simulations focusing in what happens over the mass distribution in the phase space are performed for this system. What happens through the violent relaxation period and what stands for the quasi-stationary states (QSS) of this dynamics is analyzed. The results obtained support the fact that MSP is observed already in the violent relaxation time and is maintained during the QSS. Some structures in the mass distribution function are observed. As a result of this study the mass distribution is determined by the system dynamics and is independent of the dimensionality of the system. MSP occurs in a one dimensional system as a result of the long range forces that acts in the system. In this approach MSP emerges as a dynamical feature. We also show that for HMF with different masses, the dynamical time scale is $N$.
\end{abstract}

\bibliographystyle{elsarticle-num}
\end{frontmatter}

\section{Introduction}
 Long-range interacting systems in $d$ dimensions are those described by a potential $V(r)\propto r^{-\alpha}$ with $\alpha / d< 1$. The study of these systems is challenging thermodynamics and statistical mechanics for some decades. Several well-known physical systems are long-ranged. Self-gravitating systems, non-neutral plasmas, Bose-Einstein condensates, point vortex, and others derived from those are examples of this kind of systems. These long-range interacting systems are spread all over the physics sub-areas, as astrophysics, plasma physics, condensed matter and atomic physics, hydrodynamics, nuclear physics and it defies us to study and describe them \cite{campa}. 
 
Self-gravitating systems (SGS) are of great astrophysical interest. They are ruled by the gravitational force. Dynamics of stars or galaxies clusters are examples of SGS and are objects of wide interest. The great computational effort to simulate the dynamics of SGS is a limitation, and indeed they are of considerable interest for astrophysicists. In order to study this kind of systems and to perform robust calculation of statistical physics point of view, some toy models were introduced, such as: Hamiltonian Mean Field model (HMF) \cite{hmf}, the ring model \cite{sota}, 1-d gravitating sheets model \cite{lecar}.

In this work, the HMF model will be studied with a modification from what was suggested by Antoni and Ruffo \cite{hmf} about two decades ago: the particles are non-identical, they have different masses. HMF model is a simple representative system that exhibits a long-range acting potential where every particle is subject to the effect of the mean field due to all other particles. Moreover computational calculation for the HMF model scales with the number of particles $N$, rather than $N^2$, as is the case for all the other models and realistic systems. This feature provides the possibility to perform simulations with a great number of particles and also for long evolution times.

In the very beginning of the evolution the system performs a series of rapid and strong oscillations named violent relaxation (VR), this is one feature that appears on long-ranged systems. Since 1967, D. Lynden-Bell noticed that "the violently changing gravitational field of a newly formed galaxy is effective in changing the statistics of stellar orbits" \cite{lynden-bell}. The evolution equation for the one-particle distribution function $f(\textbf{q},\textbf{p},t)$ for this collisionless system during the VR period \cite{steiner} is
	\begin{equation}
	\left(\frac{\partial}{{\partial}t} + \textbf{p}\cdot\frac{\partial}{{\partial}\textbf{q}} - \frac{{\partial}\psi}{{\partial}\textbf{q}}\cdot\frac{\partial}{{\partial}\textbf{p}} \right)f(\textbf{q},\textbf{p},t)=0,
	\end{equation}
the Vlasov equation (VE), in which $t$ is the time, $(\textbf{q},\textbf{p})$ are the conjugated phase space coordinates and $ -\partial\psi / \partial\textbf{q}$ is the mean field force $\textbf{F}(\textbf{q},\textbf{p},t)$ that acts over an individual particle.	
	
	As observed by Levin \textit{et alli} \cite{{levin},{yan}}, VR are of a macroscopic nature. These oscillations propagate as macroscopic density waves. Some particles of the system enter in resonance with these oscillations gaining energy and forming a halo distribution as a function of the collective motion. The resonant particles gain energy and reach high energy states to form a halo distribution, in the meantime these oscillations are damped and the remaining particles condense in low energy states forming a thick core. This mechanism is known as Landau damping. The final distribution achieved at the end of VR has a core-halo structure that persists during the non-equilibrium solutions of the VE, named as quasi-stationary state (QSS).

Core-halo structures are observed even for stellar or galaxy clusters. These clusters exhibit mass segregation where the lighter components are in the halo and the heavier ones are in the core \cite{alan}. This kind of mass distribution leads to mass segregation phenomena (MSP) \cite{allison1, allison2} that remains as an open problem. Actually it is not known if the MSP is dynamical or primordial.

	In next section we present the modified HMF model and the techniques used in this work. In Section 3 the results and discution will be shown. The conclusions are presented in section 4.
	
\section{Hamiltonian Mean Field with non-identical particles}
	
	Here a modification of HMF model for non-identical particles with different masses is considered. The particle mass distribution will range in the interval $[m_{min} , m_{max}]$, so that the mean field binary potential is written as
	\begin{equation}\label{hmf1}
	V_{ij} = \frac{1}{N}\sum_{i<j}^{N}\sum_{j=1}^{N} m_i m_j [1 - \cos(\theta_i - \theta_j)]
	\end{equation}
and the Hamiltonian given by
	\begin{equation}\label{eq:hamiltonian}
	H = \sum_{i = 1}^{N} \frac{p_{i}^{2}}{2m_{i}} + \frac{1}{2N}\sum_{i ,	 j = 1}^{N} m_{i} m_{j} [1 - \cos(\theta_i - \theta_j)].
	\end{equation}
	The force over the $i$-th particle is as follows
	\begin{equation}
	F_{i} = m_{i} \left[ \cos\theta_i M_{y} - \sin\theta_i M_{x} \right],
	\end{equation}
with $M$ the magnetization with components $M_x  =  \frac{1}{N}\sum_{j = 1}^{N} m_j\cos\theta_j$ and $M_y  =  \frac{1}{N}\sum_{j = 1}^{N} m_j\sin\theta_j$. In this way, the potential energy per particle written observing the Kac prescription, $v = V/N$, has the form
	\begin{equation}
	v = \frac{1}{2}\langle m \rangle^{2} - \frac{1}{2} \left[ M_{x}^{2} + M_{y}^{2} \right],
	\end{equation}
where $\langle m \rangle = \frac{1}{N}\sum_{i = 1}^{N} m_i$ is the mean mass of the system. With these equations in mind, we will present the results that we have obtained after the fulfillment of the simulations.

	The molecular dynamics simulations were performed using the fourth order symplectic integrator developed by Yoshida \cite{yoshida}. The mass range was chosen to lie in some interval. The masses were selected using the uniform random number generator \textsc{ran2} \cite{nr} in an interval $[m_{min}, m_{max}]$.

	We chose a set of water-bag ditributions as initial conditions. The water-bag distribution initial conditions was introduced by DePackh \cite{depackh}. An area in the phase space is defined and it is known to be preserved by the time evolution of the Vlasov equation. Water-bag distribution is defined as
	
	\begin{equation}
	{
		f_0(p,\theta)=\left\{
		\begin{array}{ll}
		{1}/{2p_0\hspace{1pt}\theta_0} & \mbox{if $-p_0<p<p_0$ and $0<\theta<\theta_0$,}\\
		0 & \mbox{otherwise}.
		\end{array}
		\right.}
	\label{wbinitheta}
	\end{equation}

	This function is represented by a rectangle with sides of length $2p_0$ and $\theta_0$. The particles have a uniform distribution over  positions and \textit{momenta}. 
	
	As a measure of MSP we choose two methods, both of direct observation. The first one is just the distribution of particles over the positions. The other method is the mean distances to the center of masses of the particles belonging to two categories, the particles with masses less than the mean mass $m_{L} < \langle m \rangle$, and particles with masses belonging to the higher limit $m_{H} > \langle m \rangle$, $d_L$ and $d_H$ respectively.


\section{Results and Discussions}
	In this section the results obtained by numerical simulations for the system described by Eq.(\ref{hmf1}) are presented. The inputs used on simulations appears on Fig. captions: the number $N$ of particles, the mass-range, the initial values for angle and \textit{momentum}, time step and final time. The simulation final time is beyond the VR and ends in the beginning of QSS. None of these simulations had an error over $10^{-6}$.
	
\begin{figure}[!ptb]
		\begin{center}
			\includegraphics[scale=0.35]{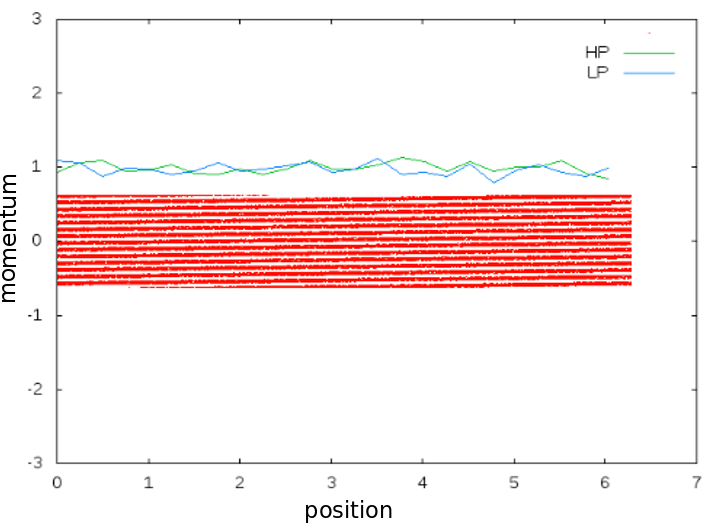}
		\end{center}
		\caption{Final phase space for a HMF with short range interaction. Simulation data: $t_f = 100$, $N = 400,000$, $\theta_0 = \pi$ and $p_0 = 1.26$ and the masses lay in this interval $m \in [1.0, 2.0]$ for all simulations. Numerical simulation time step $\Delta t = 0.1$. Here "HP" stands for \textit{ Heavy Particles} while "LP" \textit{Light Particles}. }
		\label{fig:short}
	\end{figure}

	As already mentioned, the propose of taking HMF with different masses is to see if the MSP may arise. However, before the results for the HMF with different masses are presented and in order to provide a comparation, a simulation were performed for a kind of XY model, a version of the considered HMF model in wich the interaction is due only to first neighbors. The potential considered is
	\[
	V = \sum_{i=1}^{N}m_{i} m_{i+1}[1 - cos(\theta_{i, i+1})]
	\]
where $\theta_{i, i+1} = \theta_i - \theta_{i+1}$. In  Fig. \ref{fig:short} the characteristic filamentation \cite{levin} of the phase space and the distribution of particles over $\theta$ are shown. It is clear that this system with short range interaction do not exhibit MSP.

\begin{figure}[!ptb]
		\begin{center}
			\includegraphics[scale=0.2]{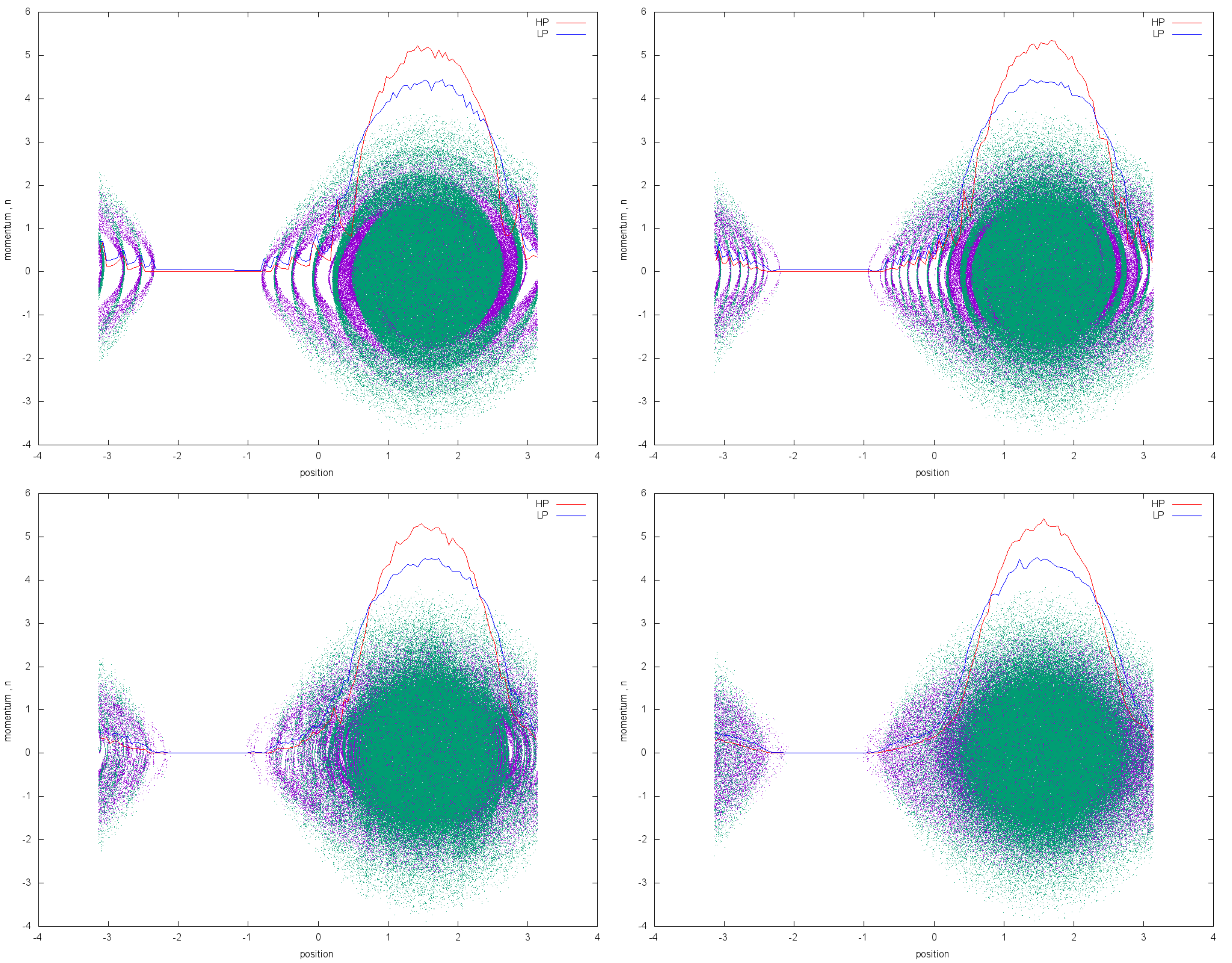}
		\end{center}
		\caption{Phase space snapshots for different instant times during the violent relaxation and QSS periods. From left to right at the top the final times are $50$, $100$ time steps and from left to right at the bottom the final times are $500$ and $5,000$. The label LP stands for light particles (represented by purple dots and the blue line, for the mass distribution), while HP for heavy particles (represented by green dots and the red line, for the mass distribution). Simulation parameters are: number of particles $N = 400,000$ with initial conditions $\theta_0 = \pi$ and $p_0 = 1.22$, which is equivalent to an energy per particle equal to $0.84$. Numerical simulation time step $\Delta t = 0.1$. }
		\label{fig:sobreposition}
	\end{figure}

	In Fig. \ref{fig:sobreposition} the first results for the system described by Eq.(\ref{eq:hamiltonian}) are shown. This Fig. displays four phase space snapshots of the system for different evolution times, $t=50, 100, 500$ and $5.000$, as stated in the Fig. caption. All the $N$ particles of the system are visible in those snapshots. This is a way to see the $N$-body behavior during the evolution. The solid lines give the mass distribution of the system in the phase portrait, red for heavy particles (HP), $m > \langle m \rangle$, and blue for the light ones (LP), $m < \langle m \rangle$. Two regions can be distinguished even in the first portrait, the core, that concentrates HP, and the halo, where the LP are more abundant. This mass distribution is maintained along the evolution. The filamentation process \cite{levin} is slightly apparent in the first and second snapshots and is seen as few layers around the core equator. The filamentation process bocomes to dissipate as a function of the time evolution of the system. In the two last snapshots, for $t=500$ and $5000$, a core-halo struture is clearly outlined. In the last snapshot only the core-halo structure survives as a consequence of the Landau damping mechanism mentioned elsewhere \cite{levin, yan}.

	In Fig. 3, four graphics showing the distribution of particles as a function of position are presented each one for a different initial energy as stated in the Fig. caption. The two graphics on the left have the initial spatial distribution laying in the $[0,\pi]$ interval and in the two graphics on the right in the $[0,2\pi]$ one. It can be seen that the two graphics on the left have similar behaviors as also the two on the right. For these two graphics on the left it is not clear that MSP appears, because the two lines purple and green are superimposed, and for the two on the right it is certain that in the core the number of HP is greater than LP and conversely in the halo the number of LP is greater than HP. For these two graphics on the right MSP is self-evident.

	\begin{figure}[!ptb]
		\begin{center}
			\includegraphics[scale=0.2]{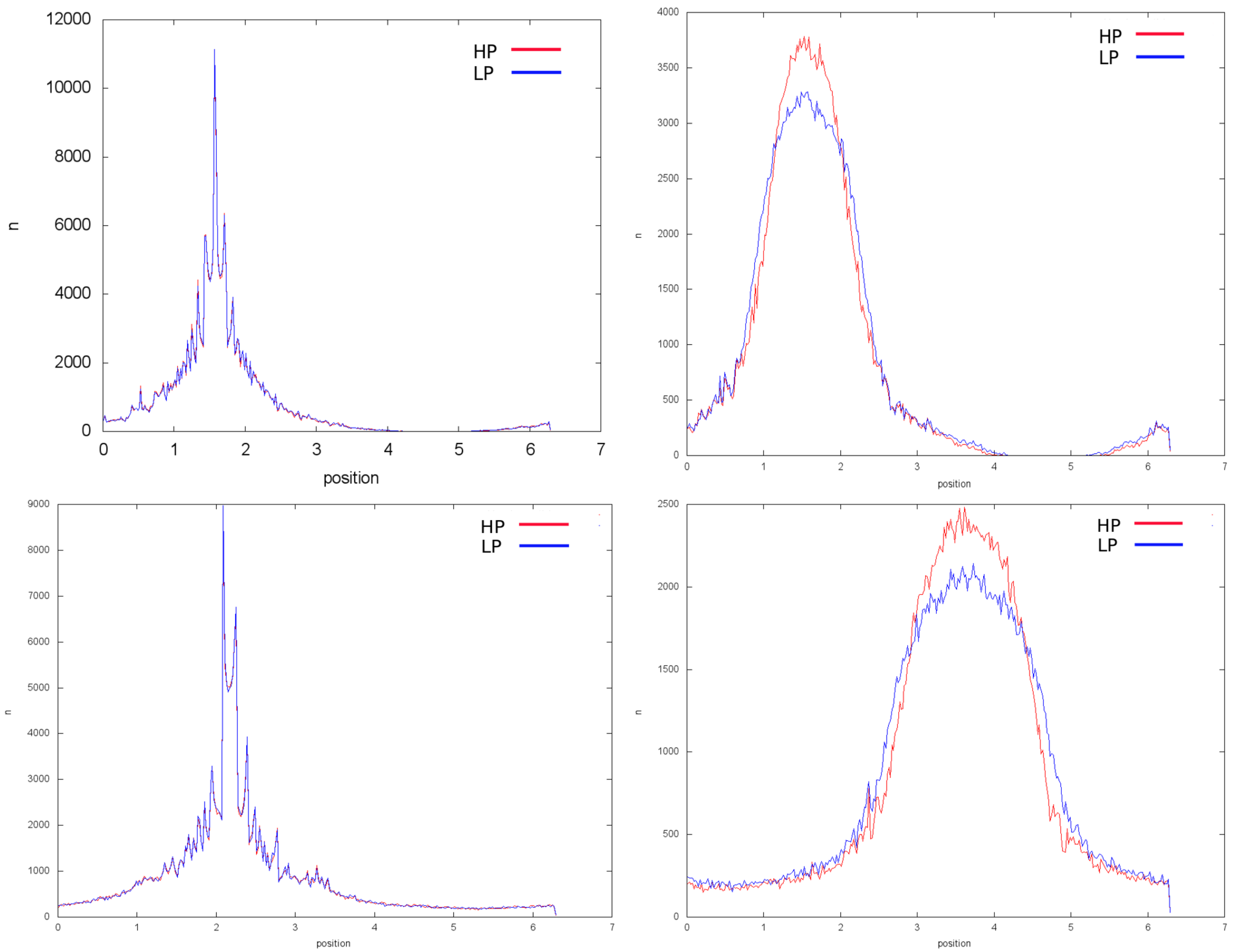}
		\end{center}
		\caption{ Comparative between particle mass distributions over positions for four different initial conditions: top left $e_0 = 0.67$, top right $e_0 = 0.85$, bottom left $e_0 = 1.12$, bottom right $e_0 = 1.30$). All simulations were performed with $N = 400,000$ and $\Delta t = 0.01$ for a final time $t=10^{4}$. The purple line stand for HP and the green one for LP. }
		\label{fig:DistsComp}
	\end{figure}

	Despite the fact that in the two graphics on the left in Fig. 3 MSP is not evident, the Landau damping process is directly shown in Fig. 4 for these set of four initial conditions. In this Fig. 3 the number of particles are expressed as function of the kinetic energy in the LP (green line) and HP (purple line) fractions. All four graphics exhibit the same information that the HP have lower kinetic energy than LP. So HP form a dense core losing energy for the LP and LP form a thin halo with energetic particles. The picture described by Levin \cite{levin} about the formation of a core-halo structure in long range interacting systems is corroborated by the results presented in Fig. 4, where LP have kinetic energy greater than HP in all situations. It can be observed that the core-halo structure leeds to mass segregation. Even for that small interval of mass variation $m \in [1.0, 2.0]$, it can be seen differences between HP and LP mass or energy distributions.

	\begin{figure}[!ptb]
		\begin{center}
			\includegraphics[angle=270,scale=0.2]{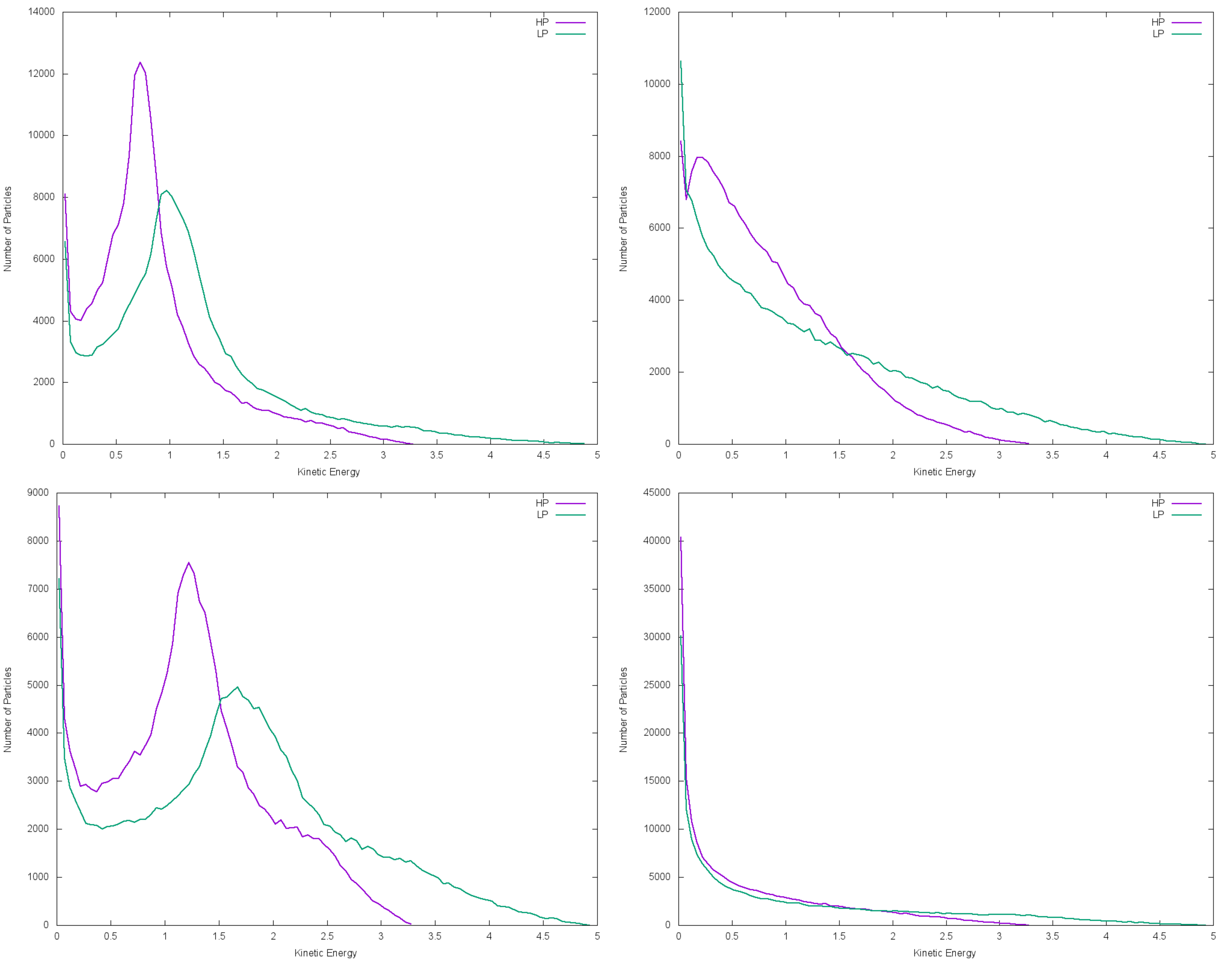}
		\end{center}
		\caption{Number of particles by kinetic energy for diferent initial energies. Top left $e_0 = 0.67$, top right $e_0=0.85$, botton left $e_0=1.12$ and botton right $e_0=1.30$. Simulation final time $T_F = 10^4$}
		\label{fig:energydist}
	\end{figure}

	Hitherto all results presented agree with each other. However if any doubt may still remain that Landau damping and MSP could not be linked, the result shown in Fig. 5 clarifies this eventual query. The quantity used to get information about MSP in a set of particles is the mean distance $d$ between these particles and their correspondent center of mass. So Fig. 5 shows a curve for the mean distance between each LP and the respective center of mass in green as a function of time and an analogous curve for HP in purple. The initial position distribution of LP and HP are identical and lay in the interval $[0,\pi]$, so for $t=0$ one can see that $d_L = d_H$. During the oscillations on the VR $d_L$ becomes different from $d_H$ and $d_L > d_H$. This fact shows that LP have a greater dispersion of its spatial distribution compared with HP that is more concentrated around its center of mass even for a simulation time of $t=40$, $d_L$ and $d_H$ reach their final values and remain stable just having small fluctuations around these values.

\begin{figure}[!ptb]
		\begin{center}
			\includegraphics[scale=0.25]{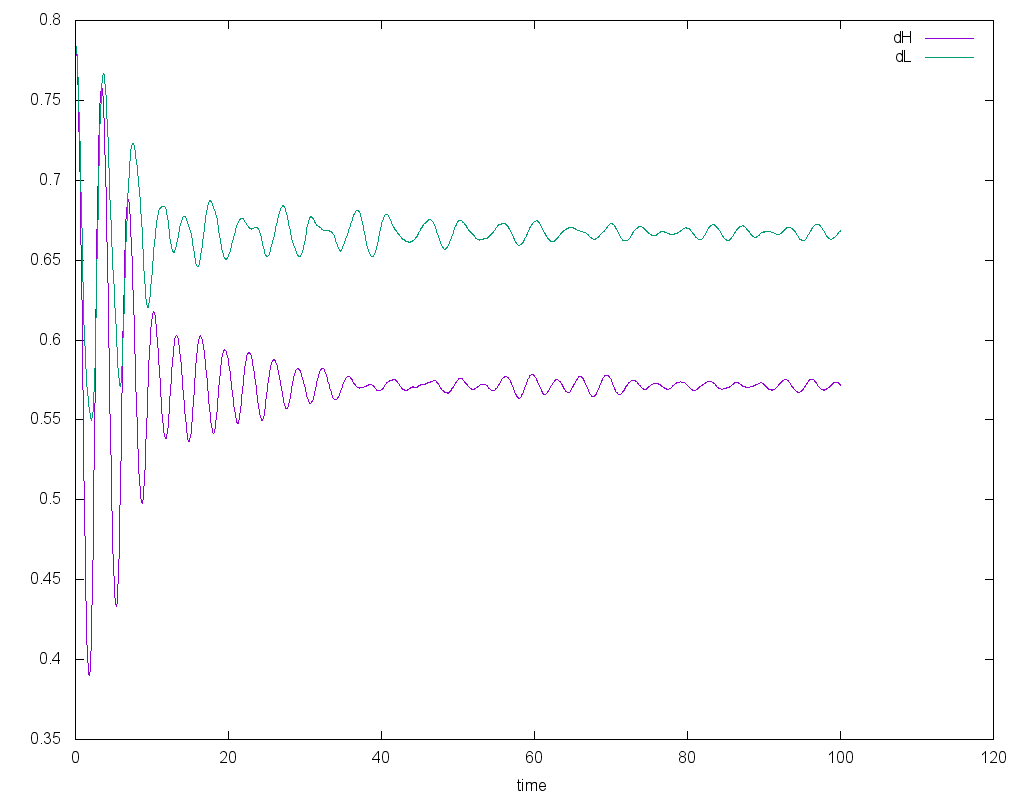}
		\end{center}
		\caption{Time evolution of $d_L$ and $d_H$. Simulation parameters: $p_0 = 1.26$, $\theta_0 = \pi$ and $N = 400,000$.}
		\label{fig:dldh}
	\end{figure}
	
	This result presented in Fig. 5 is final, MSP is a fact for HMF with different masses. And it may be repeated that each result obtained from the simulations performed in this work says that HMF presents MSP. Another feature is that the Landau damping is clearly present in HMF and its effect is MSP. The main goal here is that all results presented until now are in agreement to each other, they reinforce each other and give a robust conclusion about the existence of MSP even in an one dimensional system like HMF.

	\begin{figure}[!ptb]
		\begin{center}
			\includegraphics[scale=0.35]{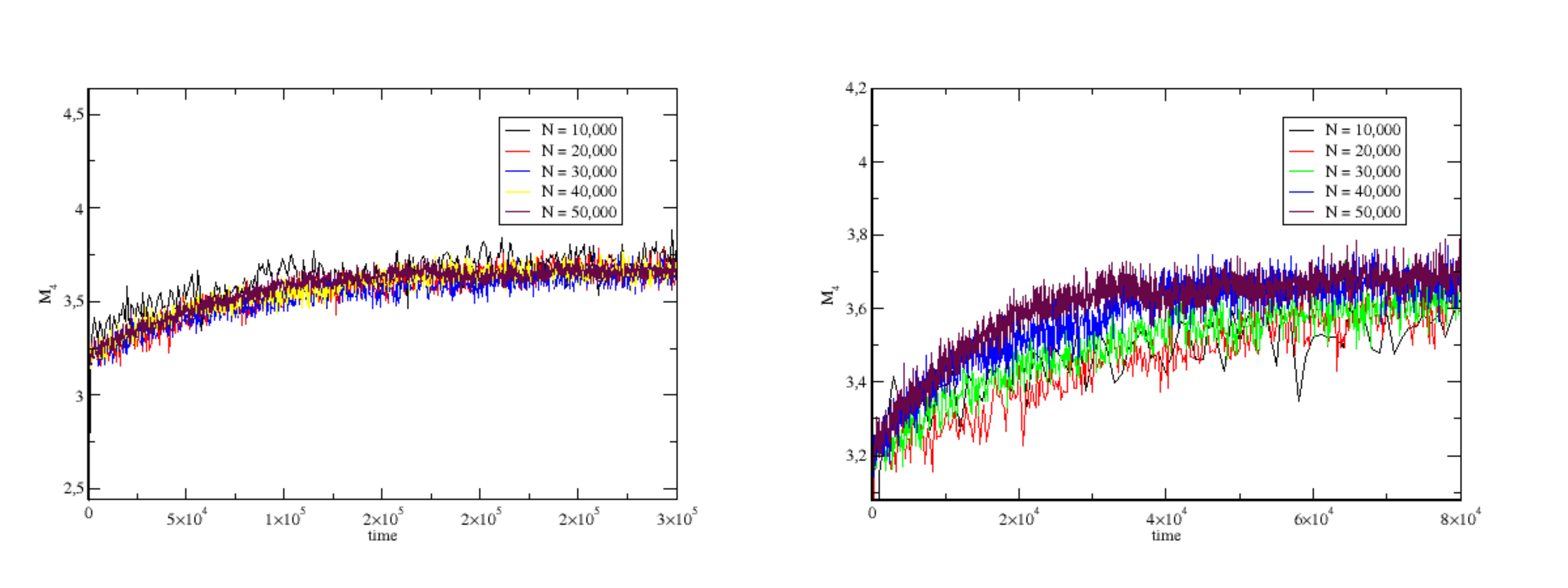}
		\end{center}
		\caption{Time scale for HMF system with different masses. Left panel there is a time rescale with $N$ while the right panel show a time rescale with $N^2$. The simulations were made to $e = 0.68$ and mass range $[1.0 , 2.0]$.}
		\label{timescale}
	\end{figure}
	
	Finally the scale of life time for the QSS comes into discussion. To do that, we present the rescaling of the forth \textit{momentum}, $M_4 = \langle p^4 \rangle$, of the dynamic \textit{momentum} distribution in Fig. 6. It is common to use the $M_4$ to show the secular evolution of the velocity (\textit{momentum}) distribution function. It is known that for the usual HMF \cite{gupta2,steiner} the $M_4$ dynamics scales with $N^2$. But comparing the two panels presented in Fig. \ref{timescale}, it can be seen that it is not the case here, where the scaling is with $N$ (left panel) instead of $N^2$ (right panel). This occurs as a consequence of the non-identical masses dynamics. We already know that \cite{steiner} for violent relaxation period the dynamics is collisionless, while for the QSS period we cannot neglect the collisional effects. It is well known that for collisions between identical particles there is just a simple exchange on momentum, evidently this is not the fact for non-identical particles and this is an important fact in trying to determine the kinetic equation for a particular system. For HMF the kinetic equation for its QSS is unknown. But the scaling property that is in discussion here have a link with the collisional integral of this unknown kinetic equation. It is evident that for HMF with different masses the $N$ dependence of the collisional integral must be implied, instead of $N^2$, as stated in \cite{steiner} where many arguments about this sort of scaling law are made.


\section{Conclusion}

	To the extent of authors' knowledge, a new version of HMF with different masses was presented in this work and maybe it was done for first hand. The main goal of doing this was to look for what happens to the mass distribution of the system after the dynamics were turned on. This new version of HMF was studied for different initial conditions and mass segregation was present for all of them. In the ambit of this HMF, MSP was found to be a dynamic feature of the system. Furthermore MSP is a consequence of the Landau damping process that is also dynamic. Landau damping was mentioned as being the cause of the core-halo structure formation in self-gravitating systems as stated by Levin \textit{et alli} in \cite{levin} and when HMF has different masses it is responsible for MSP as well.

	As a consequence of the results presented here one can note that MSP is presented in an one dimensional system. Despite the fact that MSP is common in clusters of stars and galaxies and those are three dimensional systems, it was shown that mass segregation is a phenomenon independent of the dimension of the system studied. And with this in hand it can be said that toy models like the HMF can be used to study MSP. Preliminary simulations for others well know models, like ring-model and self-gravitating sheets, also show MSP, but further studies are still on the way. 
	
	An important observation is that none of the results above can solve the question if MSP is a primordial or a dynamic phenomenon. What can be clearly said is that the dynamics leads to MSP. So the dynamic evolution drives the system over MSP once Landau damping process is turned on because of the dynamics. This is of course an insufficient answer for the question if MSP is primordial or dynamic. What one can say is that the dynamics drives the system in the direction of MSP even when the initial mass distribution is homogeneous.

	The result about the time scale for QSS is another point that deserves some attention. This is still an open problem in this field. Some previous results for different systems point out that the scaling could depend with $N^{1.7}$ \cite{yamaguchi, campa2}, while others with $e^N$ \citep{ettoumi,campa2} and some with $N^2$ \cite{gupta2, steiner}. Our simulations show that for non-identical particles, this time scales with $N$. This occurs due to the collisional regime that appear at the QSS. A solution for the problem of the time scaling of the QSS may be done following  the dynamic of correltions developed by the Bruxels' school. As it was stated it remains as an open problem of this area and a solution can be difficult to be achieved. Another problem that would be interesting to investigate in more detail is the velocity distribution shape \cite{cirto, rocha2} for light and heavy particles. This will be done in a further work.


\section{Acknowledgement}
	This work was partially supported by UESC. The authors are also thankful to professors A. Queiroz, A. Figueiredo and T. da Rocha-Filho for further discussions.

\section{References}
\bibliographystyle{plain}

\end{document}